\documentclass[11pt,a4paper]{article}
\usepackage{jheppub}

\def\h{{1 \over 2}}
\def\e{{\rm e}}
\def\al{\alpha}
\def\eps{\epsilon}
\def\D{\Delta}

\def\l{\left(}
\def\r{\right)}
\def\lsb{\left [}
\def\rsb{\right ]}

\def\si{\sin^2{\theta_W}}
\def\co{\cos^2{\theta_W}}
\def\G{\frac{g}{2\cos{\theta_W}}}
\def\g{\gamma}
\def\h{\frac12}
\newcommand{\be}{\begin{equation}}
\newcommand{\ee}{\end{equation}}
\newcommand{\bea}{\begin{eqnarray}}
\newcommand{\eea}{\end{eqnarray}}
\newcommand{\bg}{\begin{gather}}
\newcommand{\eg}{\end{gather}}
\newcommand{\bseq}{\begin{subequations}}
\newcommand{\eseq}{\end{subequations}}
\newcommand{\ov}{\overline}

\newcommand{\eqsplit}[1]{\begin{equation}\begin{split} #1\end{split}\end{equation}}

\newcommand{\nn}{\nonumber}

\title{Heavy-meson physics and
flavour violation with a single generation}
\author[a,b,c]{M.~Libanov,}
\author[a,c]{N.~Nemkov%
}
\author[a]{E.~Nugaev%
}
\author[a,b]{and I.~Timiryasov}
\affiliation[a]{
Institute for Nuclear Research of the Russian Academy of Sciences,\\
60th October Anniversary Prospect, 7a, 117312 Moscow, Russia
}
\affiliation[b]{
Physics Department, Moscow State University,\\
Vorobjevy Gory, 119991, Moscow, Russia
}
\affiliation[c]{
Department of Problems of Physics and Power Engineering,\\
Institutskii per., 9, 141700, Dolgoprudny, Moscow Region, Russia
}
\emailAdd{ml@ms2.inr.ac.ru}
\emailAdd{nemkov@inr.ac.ru}
\emailAdd{emin@ms2.inr.ac.ru}
\emailAdd{timiryasov@inr.ac.ru}
\abstract{
We study flavour-violating processes which involve heavy $B$- and $D$-mesons
and are 
mediated by Kaluza-Klein modes of gauge bosons in a previously suggested
model where three
generations of the Standard Model fermions originate from a single
generation in six dimensions. We find the bound on the size $R$ of the extra
spatial dimensions $1/R \gtrsim 3.3$~TeV, which arises from the three-body
decay $B_{s}^{0}\to K\mu e$. Due to the still too low
statistics this bound is much less stringent than
the constraint arising from $K\to \mu e$, $1/R \gtrsim 64$~TeV, which was
found in a previous work~\cite{Frere:2003ye}. Nevertheless, we argue
that a clear signature of the model would be an observation of $K\to \mu
e$ and $B_{s}^{0}\to K\mu e$ decays without observations of other
flavour and lepton  number changing processes at the same precision level.
}


\begin{document}
\maketitle

\section{Introduction}
\label{sec:Intro}

Large extra dimensions (LED) (see Ref.~\cite{Rubakov:2001kp}~for review) may
help in understanding the flavour puzzle, which is one of the most
intriguing issues of the Standard Model (SM) of particle physics. In particular, in the
previous series of works, models have been suggested~\cite{Libanov:2000uf,
Frere:2000dc} and studied~\cite{Frere:2001ug,
Libanov:2002tm,Libanov:2002ka, Frere:2003ye, Frere:2004yu,Libanov:2007zz,
Libanov:2005mv, Frere:2010ah} where a single family of fermions, with
vector-like couplings to the SM gauge fields in six dimensions, gives rise
to three generations of chiral SM fermions in four dimensions (see
Refs.~\cite{Libanov:2011st, Libanov:2011zz} for short reviews). These
generations appear as three zero modes due to a specific interaction of $6D$
fermions with fields (namely, a scalar field $\Phi$ and a $U_{g}(1)$ gauge
field\footnote{Our notations coincide with those used in
Refs.~\cite{Libanov:2000uf, Frere:2003ye}. In particular, six-dimensional
coordiantes are labeled by capital Latin indices $A,B=0,\ldots,5$.
Four-dimensional coordinates are labeled by Greek indices, $\mu ,\nu
=0,\ldots,3$. The signature is mostly negative.} $A_{A}$) which build up a
two dimensional topological defect, known as the Abrikosov--Nielsen--Olesen
vortex. The number of families $n_{f}= 3$ is however not automatically
guaranteed, but can be achieved by an adequate axial charge assignment
with respect to $U_{g}(1)$ group for the fermions. Initially the model has
been formulated in flat and infinitely large extra dimensions. Later, to
incorporate four-dimensional gauge fields, a compactified version of the
model has been developed~\cite{Frere:2003yv}. There, fermions are
localized in the core of a $(5+1)$-dimensional vortex, and two extra
dimensions form a sphere with radius $R$ accessible for (non-localized) SM
gauge bosons. Though gravity is not included in the consideration, it
should be stressed that the choice of the manifold is not important for
our principal conclusions~\cite{Frere:2003ye}. The extra dimensions can
even be infinitely large. In this case, the role of the radius $R$ of the
sphere is taken by a typical size of the localized gauge zero modes but
not by the size of the extra dimensions.

The main feature of the model which will be important in what follows is that
the three fermionic zero modes localized in the vortex background have
different generalized (supplemented by $U_{g}(1)$ global rotations) angular
momentum and, as a result, have different $\varphi $ and $\theta
$-dependencies, where $\theta $ and $\varphi $ are the polar and the
azimuthal angles on the sphere, respectively. Typically, one
has~\cite{Frere:2003yv} the following angular dependence for fermionic zero modes,
\begin{equation}
\Psi _{n}(\theta ,\varphi )\sim f_{n}(\theta ) \mathrm{e}^{i(3-n)\varphi
},\qquad n=1,2,3,
\label{Eq/Pg1/1:bmesons}
\end{equation}
where $\theta $-dependent wave functions $f_{n}(\theta )$ behave near
the origin, as:
\begin{equation}
f_{n}(\theta )\sim \theta ^{3-n},\qquad \theta \to 0\;
\label{Eq/Pg2/1:bmesons}
\end{equation}
($\theta=0$ corresponds to the center of the vortex).

If the Brout-Englert-Higgs (BEH) scalar $H$ couples to the defect, its
classical $\varphi $-independent configuration can
also be nonzero in the core with a typical size $R\theta _{H}\simeq
R\theta _{\Phi }$~\cite{Libanov:2007zz, Libanov:2005mv}, where $R\theta
_{\Phi }$ is a typical size of the scalar $\Phi $.
With different wave function profiles (\ref{Eq/Pg1/1:bmesons}),
(\ref{Eq/Pg2/1:bmesons}) for different modes, their overlap with  $H$
leads to a power-like hierarchical structure of masses and mixings from
the 4-dimensional point of view
\begin{equation}
m_{33}:m_{22}:m_{11}\sim 1:\delta ^{2}:\delta ^{4}\;,
\label{Eq/Pg2/2:bmesons}
\end{equation}
which is governed by a small parameter~\cite{Frere:2003yv}
\begin{equation}
\delta =\frac{\theta _{\Phi }}{\theta _{A}}\sim
\sqrt[4]{\frac{m_{d}}{m_{b}}}\sim 0.1\;,
\label{Eq/Pg2/3:bmesons}
\end{equation}
where $\theta _{A}\simeq 0.1$ is the angular size of the vortex gauge field
$A_{A}$. Thus, one sees from Eqs.~(\ref{Eq/Pg2/1:bmesons}),
(\ref{Eq/Pg2/2:bmesons}) that number $G$, which enumerates the 4-dimensional 
fermionic generations, is
nothing but the angular momentum $n$ of the zero modes from the 6-dimensional
point of view. It is worth noting that in the absence of mixings which
appear from the fermion couplings with $\varphi $-dependent $\Phi $ and
$\varphi $-independent $H$, the angular momentum, and therefore the generation
number, are strictly conserved quantities.

The Kaluza-Klein (KK) spectrum of the SM gauge bosons breaks up into two
groups of the modes~\cite{Frere:2003ye}. The first group contains modes
which do not depend on $\varphi $, and so have zero angular momentum. In
particular, zero modes of the gauge bosons, which correspond to the usual
$4D$ SM gauge fields, are independent from extra dimensional coordinates
$\varphi $ and $\theta $, and therefore belong to this group. The second
group contains $\varphi $-dependent modes only. The $4D$ mass spectrum of
these fields starts from $\sqrt{2}/R$. Since the fields of the second group
are $\varphi $-dependent, they carry angular momentum and so generation
number. As a result, they can (and have to) violate flavour and/or lepton
number. Indeed, from the 6-dimensional point of view, say, $\mu $ and $e$
are modes of the same fermion, the difference is only in their $6D$ angular
momenta. Therefore, the angular momentum of the state $\mu ^{+}e^{-}$ is
equal to 1 ($2$ for $e^{-}$ and $-1$ for $\mu ^{+}$). The same is true for
the state $d\bar{s}$. Since both states interact with $6D$ photon and
$Z$-boson there are KK modes of the gauge bosons which carry unit angular
momentum and lead to the forbidden decay $K^{0}\to \mu e$. The decay rate
of this process is suppressed by the masses of the KK modes, that is, by
$R^{4}$. Note that this process does not violate generation number but violates
lepton number and takes place even in the absence of the intergeneration
mixings. If one takes into account mixings then other FCNC processes
become allowed, e.g. $\mu \to 3e$ $(\Delta G=1)$ or additional
contributions to $CP$ violation in kaons  and kaons mass difference
$(\Delta G=2)$. However, the amplitudes of these processes in addition to
the suppression by the masses of KK modes are suppressed at least by the
factor $(\epsilon \delta )^{|\Delta G|}$ where $\epsilon \sim 0.1\div 1$
is the parameter, which governs mixings. Thus, the latter processes are
typically less restrictive than the flavour-conserving, but lepton number
violating processes.

The specific pattern of these flavour(lepton number)-violating effects, which
could distinguish the models of this class from other LED models by
signatures in rare processes at low energy, has been studied in
Ref.~\cite{Frere:2003ye}. In particular, the forbidden
kaon decays $K^{0}\to \mu e$, $K^{+}\to \pi ^{+}\mu ^{+}e^{-}$ ($\Delta
G=0$), lepton flavour violation $\mu \to 3e$, $\mu \to e\gamma $, $\mu
e$-conversion ($\Delta G=1$), and additional contributions to $K_{L}-K_{S}$
mass difference and $CP$-violation in kaons ($\Delta G=2$) have been
studied. It has been found that the strongest constraint on the size of
the extra-dimensional sphere (or the size of the gauge bosons localization)
$R$,
\begin{equation}
\frac{1}{R} \gtrsim \frac{1}{R_{K\to \mu e}}\equiv 64 \mbox{ TeV}\;,
\label{Eq/Pg3/1:bmesons}
\end{equation}
arises from non-observation of the decay $K^{0}\to \mu e$. A clear
signature of the model would be an observations of $K^{0}\to \mu e$ decay
without observation of $\mu \to 3e$, $\mu \to e\gamma $ and $\mu
e$-conversion at the same precision level.

On the other hand, recently there has been significant progress in
studying of the physics of heavy $B$ and $D$-mesons mostly due to the CLEO-c ~\cite{Gao:2004xg},
Belle, BaBar ~\cite{Miyazaki:2010zz, Trabelsi:2010zz} and LHCb~\cite{Meadows:2011bk} experiments. Therefore, the
main goal of this paper is to study  phenomenological constraints on the
extra-dimensional size $R$ arising from an analysis of rare processes in
the heavy mesons physics. Our aim is twofold. First, we obtain the
constraints on $R$ from the present-day experimental data.
Second, we single out those processes with the heavy mesons, which have
maximal probability if the radius $R=(64\mbox{ TeV})^{-1}$.

The paper is organized as follows. In
Sec.~\ref{Section/Pg4/1:bmesons/Effective four-dimensional Lagrangian} we
give a brief description of the Lagrangian responsible for the
flavour-violating processes. We study specific flavour-changing effects
involving $B$-mesons in Sec.~\ref{Section/Pg7/1:bmesons/Flavour violating
processes in B-physics} and $D$-mesons in
Sec.~\ref{Section/Pg16/1:bmesons/Constraints from processes with
D-mesons}. We conclude in Sec.~\ref{Section/Pg17/1:bmesons/Conclusions}
with a description of distinctive features specific for the given class of
models.

\section{Effective four-dimensional Lagrangian}
\label{Section/Pg4/1:bmesons/Effective four-dimensional Lagrangian}

The effective 4$D$ Lagrangian, which is responsible for the
flavour-violating effects, has been obtained in Ref.~\cite{Frere:2003ye}.
Here we give some results mostly with the aim to introduce notations,
which are used in what follows.

The 4-dimensional interaction of the fermionic zero modes with KK tower of
the photon is given by (all fields depend on 4-dimensional coordiantes
only)
\be
{\cal L}_4=e\cdot{\rm Tr}({\bf A}^\mu {\bf  j^*}_\mu ),
\label{L}
\ee
where $e$  is the usual 4-dimensional electric charge,
\be
{\bf A}^\mu =({\bf A}^\mu )^\dag =\sum \limits_{l=0}^{\infty } \left(
\begin{array}{ccc}
E_{11}^{l,0}A_{l,0}^\mu &E_{12}^{l,1}A_{l,1}^\mu & E_{13}^{l,2}A_{l,2}^\mu \\
E_{21}^{l,1}A_{l,1}^{\mu*}&E_{22}^{l,0}A_{l,0}^\mu &
E_{23}^{l,1}A_{l,1}^\mu \\
E_{31}^{l,2}A_{l,2}^{\mu*}&E_{32}^{l,1}A_{l,1}^{\mu*}& E_{33}^{l,0}
A_{l,0}^\mu,
\end{array}
\right),
\label{Agauge}
\ee
and
\be
j_{m n}^\mu=a^\dag_m\bar{\sigma}^\mu a_n,
\ee
where $a_n$ are two-component Weyl spinors. Indices $m, n$ enumerate
generation number, and overlap constants $E_{mn}^{l,n-m}$ can be estimated
as
\begin{equation}
E_{mn}^{l,m-n}\simeq \left\{
\begin{array}{lcl}
\displaystyle l^{|m-n|+1/2}\theta _A^{|m-n|} & \mbox{at} &
l\theta_A\ll1,\\
\displaystyle \frac{1}{\sqrt{\theta _A}} & \mbox{at} &
l_{\mbox{max}}\simeq\displaystyle\frac{1}{\theta _A},\\
\displaystyle {\rm e}^{-lF(\theta _A)}       & \mbox{at} &
l\theta_A\gg1.
\end{array}
\right.
\label{E}
\end{equation}

The fields $A_{l, m}^\mu (x)$ are  expansion coefficients of 6-dimensional
field $\mathcal{ A}^A (x,\theta ,\varphi )$ in spherical harmonics $Y_{l,
m}(\theta, \phi)$. The subscript $l$ in (\ref{Agauge}) enumerates modes with
$4D$ masses  $m^{2}_{l}=l(l+1)/R^{2}$, while the subscript $m$ corresponds
to the value of the carried 6D angular momentum or, what is the same,
to the generation number $G$.

We see that fermions have the strongest couplings to the heavy modes
with masses
\[
m_l=\frac{\sqrt{l(l+1)}}{R}\sim \frac{1}{\theta _AR}.
\]
The reason for this is obvious: modes with $l\sim 1/\theta$ have the largest
overlaps with fermionic wavefunctions of the size $\theta$
($\theta\approx\theta_A$ in our case);
(lower
modes have larger width in $\theta$ while higher modes oscillate several
times at the width of the fermions). We stress that this feature depends
neither on details of localization of fermions and gauge bosons nor on the
shape and size of extra dimensions.

It is worth noting that $4D$ scalars $\mathcal{ A_{\theta }}$ and
$\mathcal{ A}_{\varphi }$ do not interact with the fermionic zero modes
and so we omit them in what follows.

The matrix elements of (\ref{Agauge}) and the fermions $a_{n}$, which enter
the Lagrangian (\ref{L}), are the states in the gauge basis, while
physically observed mass eigenstates are their linear
combinations. In particular, the mass matrix of the fermions with quantum
numbers of the down-type quarks is given \cite{Frere:2000dc,Frere:2003yv}
by
\begin{equation}
M_D= \left(
\begin{array}{ccc}
m_{11}&m_{12}&0\\
0&m_{22}&m_{23}\\
0&0&m_{33}
\end{array}
\right) \propto \left(
\begin{array}{ccc}
\delta ^4&\epsilon \delta ^3&0\\
0&\delta ^2&\epsilon \delta \\
0&0&1
\end{array}
\right),
\nonumber
\end{equation}
where $\delta $ is given in (\ref{Eq/Pg2/3:bmesons}) and
$\epsilon \sim 0.1$ is the parameter, which governs mixings.
To diagonalize the mass matrix one should use biunitary
transformations,
\[
S^\dag _dM_DT_d=M_D^{\rm diag}.
\]

The  fermions in the mass basis are
\[
Q_n=(S^\dag _d)_{nm}q_m\;,\ \ \ D_n=(T^\dag _d)_{nm}d_m,
\]
where we denoted $a_n$ as $q_n$ for the left-handed and as $d_n$ for
the right-handed down-type quarks. If one rewrites the current {\bf
j}$_\mu$ in terms of the mass eigenstates, then the matrix {\bf A}$^\mu$,
Eq.~(\ref{Agauge}), should be replaced by
\[
\tilde{\bf A}^\mu= S^\dag _d{\bf A}^\mu S_d.
\]
and to the leading order in $\alpha,\gamma $ and $\epsilon
$ takes the form
\begin{equation}
\left(\! \! \!
\begin{array}{ccc}
{\bf A}_{11}-2{\rm Re}(\epsilon^* \alpha {\bf A}_{12})&{\bf
A}_{12}\! +\! \epsilon \alpha ({\bf A}_{11}\! \!  -\! \! {\bf A}_{22}\!
)\! -\! \gamma \epsilon^* {\bf A}_{13} &{\bf A}_{13}+\epsilon (\gamma {\bf
A}_{12}-\alpha {\bf A}_{23})\\
\\
\! {\bf A}_{12}^*\! \! +\! \epsilon^* \alpha ({\bf A}_{11}\! \! -\!
\! {\bf A}_{22})\! -\! \epsilon \gamma {\bf A}_{13}^* & {\bf A}_{22}\!
+\! \displaystyle 2 {\rm Re}(\epsilon^*\! (\alpha {\bf A}_{12}\! -\!
\gamma {\bf A}_{23})) &{\bf A}_{23}\!\!  +\! \epsilon^* \alpha
{\bf A}_{13}\! \! +\! \! \epsilon \gamma ({\bf A}_{22}\! \! -\! \! {\bf
A}_{33})\\
\\
{\bf A}_{13}^*+\epsilon^*(\gamma {\bf A}_{12}^*-\alpha {\bf A}_{23}^*) &
{\bf A}_{23}^*\! +\! \epsilon \alpha {\bf A}_{13}^*\!\!  +\epsilon ^*\! \!
\gamma ({\bf A}_{22} \! \! -\! \! {\bf A}_{33})& {\bf
A}_{33}+2{\rm Re}(\epsilon^* \gamma {\bf A}_{23})
\end{array}
\! \! \! \right)^{\displaystyle\mu}\;\!\!\!\!\!\!,
\label{A matrix}
\end{equation}
where the parameters $\alpha\simeq\delta$ and $\gamma\simeq\delta$ are dimensionless 
combinations of the matrix elements of $M_D$ (see Ref.~\cite{Frere:2003ye} for details).

The interaction of fermions with $W^\pm$ and $Z$ bosons is very similar to
the electromagnetic couplings discussed above. There are two differences:
firstly, the current  ${\bf j}_\mu$ in Eq.~(\ref{L}) is
replaced by the SM charged and neutral weak currents;
secondly, the gauge eigensystem is modified as discussed in
Ref.~\cite{Frere:2003ye}. The latter modification does not change
the results significantly: it is negligible for KK modes and it
does not result in flavour violation for the lowest mode.

As an example, let us discuss interactions of the $Z$-boson KK tower,
which contribute to the flavour(lepton number)-violating processes. To be
specific, let us consider the $\{23\}$ element of the matrix (\ref{A matrix}),
${\bf Z}_{23}^\mu+\epsilon^* \alpha {\bf Z}_{13}^\mu+\epsilon \gamma ({\bf
Z}_{22}^\mu-{\bf Z}^\mu_{33})$. It indicates that the interaction of
the fermionic current $j^\mu_{2 3}$  with gauge bosons $Z_{l, 1}^\mu$
is unsuppressed, interaction with $Z_{l, 2}^\mu$ is suppressed by
$(\epsilon^*\alpha)$ and interaction with $Z_{l, 0}^\mu$ is suppressed
(according to \eqref{E}) by $\epsilon\gamma \l E_{2 2}^{l, 0} -
E_{33}^{l, 0} \r$. Taking into account that the subscripts $(m n)$ in $j_{m
n}^\mu$ denote generation numbers one  obtains the leading
(unsuppresed) interactions of neutral currents,
\be
{\cal L}_{NC}=\G\sum_{l=1}^\infty E^{l,1}_{23}Z^\mu_{l,1}\l g_R^q \bar{s}
O_\mu^R b + g_L^q \bar{s} O_\mu^L b + g_R^l \bar{\mu  } O_\mu^R \tau  +
g_L^l \bar{\mu  } O_\mu^L \tau  \r + \mbox{ h.c.}\;,
\ee
where $O^{L,R}_\mu = \frac{1\pm\gamma_5}{2}\gamma_\mu$ and constants
$g_{L, R}$  defined in the usual way,
\begin{align*}
g_L^q& = - \h + \frac13 \sin^2\theta_W, &  g_R^q& = \frac13 \sin^2\theta_W,\\
g_L^l& = - \h + \sin^2\theta_W, &  g_R^l& = \sin^2\theta_W.
\end{align*}

To confront the model with the experimental results, one needs to
integrate out the heavy KK modes and calculate the effective four-fermion
coupling $g_{mn}$, that is, in each particular case, to sum up the
contributions
\[
g_{mn}^l=e^2\frac{(E_{mn}^{l,m-n})^2}{m_l^2}
\]
for all $l$. A very naive estimate gives, using Eq.~(\ref{E}),
\begin{equation}
g_{mn}\sim e^2l_{\mbox{max}}\cdot R^2\theta _A=\frac{e^2}{\theta _A}\cdot
R^2\theta _A=e^2R^2
\nonumber
\end{equation}
so that
\[
\frac{g_{mn}}{G_F}\sim(M_WR)^2.
\]

\section{Flavour violating processes in B-physics}
\label{Section/Pg7/1:bmesons/Flavour violating processes in B-physics}


Our purpose is to determine which processes involving $B$-mesons are the
most sensitive to the new physics in the context of the model under
consideration.

In the case of $K$-mesons  $K\to \mu e$ decay appeared to be the most
restrictive \cite{Frere:2003ye}. This is a special feature of the theory:
usually in the frameworks of models with LED the kaon mass difference and
$CP$-violation in kaons are the most sensitive to the physics beyond SM
(see, e.g., ~\cite{Delgado:1999sv}). The reason is that $\D m_K$ arises
from transition $K^0\leftrightarrow \overline{K}^0$ changing generation
number to $\D G=2$. But in our case the corresponding contribution is
suppressed by factor $(\delta \eps)^2$. This is also the case for
$B$-mesons. Processes with $\D G=1$, e.g. $B^0\to \mu e$ or $B^0_s \to
\mu^+\mu^- $, and $\D G=2$ transitions, responsible for $B^0_s - \bar{B}^0_s$ mass
difference, are suppressed in the same way.

Taking into account this observation, among large number of
possible flavour-violating processes we single out, on the one hand,
those with the most stringent experimental bounds, and, on the other hand,
those whose  amplitudes are suppressed less by the mixings. These
processes are collected in Table~\ref{table} where we have used
data from~\cite{Nakamura:2010zzi}. Below we present detailed
calculations for both $\D G=0$ and $\D G\neq 0$ processes.

\begin{table}[t]
\begin{center}
\begin{tabular}{|l|c|c|}
\hline Process & Branching ratio (BR)  & $|\D G|$\\
\hline $B^0_s\to \mu e$&$<2.0 \cdot 10^{-7}$&0\\
\hline $B^0\to \tau e$&$<2.8 \cdot 10^{-5}$& 0\\
\hline $B^0\to K^0\mu e$&$<2.7 \cdot 10^{-7}$&0\\
\hline $B^0\to \mu e$&$<6.4 \cdot 10^{-8}$&1\\
\hline $B^0_s\to\mu^+\mu^-$& $<4.7\cdot 10^{-8}$&1\\
\hline $B^0_s\leftrightarrow \ov{B}^0_s$& $\Delta m_{B^0_s}\approx
1.17\cdot 10^{-8}$ MeV & 2\\
\hline
\end{tabular}
\end{center}
\caption{List of flavour(lepton number)-violating processes involving
$B$-mesons which are most sensitive to new physics in the context of the
model.
\label{table}}
\end{table}


\subsection{Processes with $\D G= 0$}
\label{Subsec/Pg9/1:bmesons/Processes with D G=0}

\subsubsection{$B^0_s\to \mu e$}
Let us consider $B^0_s\to \mu e$ decay which violates lepton family
number. Its branching ratio (BR) is  most strongly bounded
among two-body $B^0_s$ decays with $\D G=0$. Under experimental
circumstances one does not distinct $B^0_s$ or $\bar{B^0_s}$ in the initial
state and $\mu^+ e^-$ or $\mu^- e^+$ in the final state. Thus, four processes
possibly can contribute to experimental BR. However two of them $B^0_s\to
\mu^+ e^-$ and $\bar{B^0_s}\to \mu^- e^+$ correspond to $\Delta G=0$ while
two others to $\Delta G=2$ and therefore are sufficiently suppressed.
Therefore, we  calculate the width of $\bar{B^0_s}\to \mu^-e^+$ decay and
take into account that
$\Gamma(B^0_s\to\mu^+e^-)=\Gamma(\bar{B^0_s}\to\mu^-e^+)$.

Since $B^0_s$ is a pseudoscalar it cannot decay through purely vector
interaction of the KK modes of the photon. However, the higher modes of the
$Z$ boson interact with a $V - A$ current and contribute to the decay
width. Dominant  axial coupling in the effective four-dimensional
Lagrangian has form,
\begin{equation}
\G \sum_{l=1}^\infty
Z^\mu_{l,1}\left\{E^{l,1}_{23}\bar{s}\gamma_\mu(-\frac12\gamma_5)b+
E^{l,1}_{12}\bar{e}\gamma_\mu(2\si-
\frac12-\frac12\gamma_5)\mu\right\}\;, \nn
\end{equation}
where constants $E^{l,m-n}_{m n}$ characterise overlap of the $Z$-boson KK
modes with the fermionic wave functions, see \eqref{E}.

To obtain the effective four-fermion coupling one has to sum over all
intermediate KK modes in a way similar to
Sec.\ref{Section/Pg4/1:bmesons/Effective four-dimensional Lagrangian},
\be
G_{eff}=\l\G\r^2\sum_{l=1}^\infty
\frac{E^{l,1}_{23}E^{l,1}_{12}R^2}{l(l+1)}=\l\G\r^2\zeta R^2\;, \nn
\ee
where $l(l+1)/R^2$ is the mass squared of the $l$-th gauge boson mode and
$\zeta\simeq0.47$ is the result of numerical evaluation of the sum.

Now it is straightforward to write down the amplitude for $B^0_s\to \mu e$
decay,
\begin{equation}
 M=\l\G\r^2\zeta R^2 L_\mu H^\mu\;,
\label{M}
\end{equation}
where ${L_\mu=\bar{e}\gamma_\mu(-\frac12\gamma_5-[\frac12-2\si])\mu}$ is
the leptonic current, $H^\mu=-\frac12 f_{B_s} p^\mu \phi_B$ is the
hadronic current;  $p^\mu$, $f_{B_s}$ and $\phi_B$ are momentum, decay
constant and the wave function of  $B_s$-meson, correspondingly. We use
the numerical value $f_{B_s}\simeq 200 \mbox{~MeV}$ from \cite{Collins:2000ix}.

From Eq. (\ref{M}) one obtains the partial decay width,
 \begin{equation}
 \Gamma(B^0_s\to\mu^+e^-)=\frac{G^2_F m^4_Z\zeta^2 R^4 f^2_{B_s}
 m_{B_s}m^2_\mu (1+(1-4\si)^2)}{128\pi}.
\label{Eq/Pg10/1:bmesons}
 \end{equation}
Using the experimental limit from Table~\ref{table},
\begin{equation}
Br(B^0_s\to \mu
 e)=2\Gamma(B^0_s\to\mu^+e^-)\cdot\tau_{B_s^0} < B_{B^0_s\to \mu
 e}=2.0\cdot 10^{-7}\;,
 \label{Eq/Pg10/2:bmesons}
\end{equation}
where $\tau _{B_{s}^{0}}$ is the $B$-meson lifetime,
we obtain the constraint on the size of the sphere $R$,
 \begin{equation}
 \frac1R>m_Z\left(\frac{G^2_F\zeta^2 f^2_{B_s} m_{B_s}m^2_\mu
\tau _{B_{s}^{0}} (1+(1-4\si)^2)}{64\pi B_{B^0_s\to \mu e}}\right)^{1/4}\;.
 \label{Bmue}
 \end{equation}
 Substituting here all necessary numerical values from
 Ref.~\cite{Nakamura:2010zzi} we find,
 \begin{equation*}
 \frac1R > 0.7 \mbox{~TeV.}
 \end{equation*}

On the other hand, one can use the constraint, arising from the rare kaon
decay (\ref{Eq/Pg3/1:bmesons}), to find out what is BR of the $B$-meson if
the size of the sphere satisfies to (\ref{Eq/Pg3/1:bmesons}).
Substituting Eqs. (\ref{Eq/Pg10/1:bmesons}), (\ref{Eq/Pg3/1:bmesons}) into
(\ref{Eq/Pg10/2:bmesons}), we obtain
\begin{equation}
 Br(B^0_s\to \mu e)|_{R<R_{K\to \mu e}}<(64\pi)^{-1} m_Z^4
R_{K\to \mu e}^4G^2_F\zeta^2 f^2_{B_s} m_{B_s}m^2_\mu\tau _{B_{s}^{0}}
(1+(1-4\si)^2)\;,\nn
\end{equation}
or numerically,
\begin{equation}
Br(B^0_s\to \mu e)|_{R<R_{K\to \mu e}}<4.2 \cdot 10^{-15}\;.
\label{Eq/Pg10/3:bmesons}
\end{equation}

 \subsubsection{$B^0\to \tau e$}

 $B^0\to \tau e$ decay has two order larger experimental constraint on
BR than $B^0_s\to \mu e$. However, since $m_\tau/m_\mu\simeq 17$ we can
expect a comparable result. Consideration of these two processes is quite
similar from both experimental and theoretical points of view.

 Interaction vertex obtained from \eqref{A matrix} is
 \begin{equation}
 \G\sum_{l=2}^\infty
 Z^\mu_{l,2}E^{l,2}_{13}\left\{\bar{b}
\gamma_\mu(-\frac12\gamma_5)d+\bar{e}\gamma_\mu(2\si-
 \frac12-\frac12\gamma_5)\tau\right\}\;,\nn
 \end{equation}
 and after obvious replacements, one obtains from
(\ref{Bmue}),
\begin{equation}
 \frac1R>m_Z\left(\frac{G^2_F\xi^2 f^2_{B^0}
m_{B^0}m^2_\tau\tau_{B_{d}^{0}} (1+(1-4\si)^2)}{64\pi B_{B^0\to \tau
e}}\right)^{1/4}\nn
\end{equation}
where
\[
\xi=\sum^\infty_{l=2}\frac{(E^{l,2}_{13})^2}{l(l+1)}\simeq 0.27\;,
\]
$\tau _{B_{d}^{0}}$ is the $B_{d}^{0}$-meson lifetime, and $B_{B^0\to
\tau e} = 2.8\cdot 10^{-5}$ is the experimental bound on $B^0\to \tau
e$ BR. Numerically,
\begin{equation}
 \frac1R > 0.65 \mbox{~TeV}.\nn
 \end{equation}
On the other hand,
\be
Br(B^0\to \tau e)|_{R<R_{K\to \mu e}}<4.1\cdot 10^{-13}\;,\nn
\ee
that is two order of magnitude larger than (\ref{Eq/Pg10/3:bmesons}), which
is a consequence of the large ratio $m_{\tau }/m_{\mu }$.

\subsubsection{$B^0\to K^0\mu e$}

This decay has the stringent experimental bound on BR among three-body decays with $\D
G=0$.

Vertex, responsible for the $bs\rightarrow\mu e$ transition, is
\eqsplit{ &e\sum_{l=1}^\infty A_{l,1}^\mu \l -E_{2 3}^{l,1}{1\over 3}
\ov{b}\g_\mu s + E_{1 2}^{l,1}\ov{e}\g_\mu \mu \r + \\
& \G\sum_{l=1}^\infty  Z_{l,1}^\mu \lsb E_{2 3}^{l,1}\l {2\over 3}\si-
\frac12 \r \ov{b}\g_\mu s -E_{1 2}^{l,1}\ov{e}\l 2\si-\frac12+\frac12\g_5
\r \g_\mu \mu \rsb \nn }
It is worth noting that, because both $K$ and $B$ are pseudoscalars,
the process is now mediated by both $Z$ and photon modes.

The matrix element of hadronic current between
external meson states is parametrized as follows (see
\cite{Colangelo:1995jv}):
\be
\langle K(p'')|\ov{s}\g_\mu b |B(p')\rangle =(p'+p'')_\mu
F_1(q^2)+\frac{M_B^2-M_K^2}{q^2}q_\mu \l F_0(q^2)-F_1(q^2) \r , \nn
\ee
where $q=p'-p''$ is the total momentum of the leptons.

In the limit of the vanishing lepton masses the only formfactor $F_{1}$
gives contribution to the partial decay width. In the kinematically
allowed region $F_1$ is given by,
\be
 F_1(q^2)=\frac{F_1(0)}{1-\frac{q^2}{M^2}}\;,
\nonumber
 \ee
 with $F_1(0)\simeq0.25$ and $M\simeq M_B$.
The amplitude of the decay in this limit  is
\begin{equation}
M  = 2g^2 \zeta R^2 p'_\mu  F_1(q^2)\ov{e} \g^\mu \l C_V-C_A \g_5 \r
\mu\;,
\label{Eq/Pg12/1:bmesons}
\end{equation}
with
\eqsplit{ C_V&={1\over 3}\si-{1\over 4\co}\l \h - {2\over 3}\si \r \l \h - 2\si \r,\\
C_A&=\h\cdot {1\over 4\co}\l \h - {2\over 3}\si \r . \nn}
Calculating the partial decay width from
(\ref{Eq/Pg12/1:bmesons}), one finally obtains,
\begin{equation}
{1\over
R}>m_W\left(\frac{G^2_F\zeta^2F^2_1(0)m^5_{B^0}
\tau_{B^0}(C^2_V+C^2_A)} {6\pi^3 B_{B^{0}\to K\mu
 e}}\right)^{1/4}\;.
\nonumber
\end{equation}
Using the value $B_{B^{0}\to K\mu e}=2.7\cdot 10^{-7}$ for the
experimental constraint, we set the limit,
\begin{equation}
\frac1R > 3.3 \mbox{~TeV}\;, \nn
\end{equation}
and for the bound on $R$ obtained from $K\to \mu e$ decay, BR for considering process would be:
\be
Br(B^{0}\to K\mu e) \left |_{R<R_{K\to \mu e}}\right .<2.4\cdot 10^{-12}\;. \nn
\ee
Note, that this bound for BR is the same as for $K\to \mu e$ decay at $1/R=64\mbox{~TeV}$.

\subsubsection{$\D G=0$: Summary}
To summarize this subsection, among B-mesons decays with $\D G=0$ the most
rigid restriction has been obtained from the three-body decay while
the constraint, arising from the two-body decay with the same BR
($B^0_s\to \mu e$) is the five times smaller. This result is in contrast
to the constraints coming from the kaon decays, where two-body decay $K\to\mu e$
gives the best result with compare to three-body decay $K^+\to\pi^0\mu^+\nu$ \cite{Frere:2003ye}.
 To clarify this let us compare results for three-
and two-body decays,
\be
\left .\frac{(1/R)_{3}}{(1/R)_{2}}\right |_i=N\l
F_i^2\frac{Br_{3\,i}}{Br_{2\,i}}\l\frac{m_i}{f_i}\r^2 \l
\frac{m_i}{m_\mu}\r^2\r^{1/4}\;, \nn
\ee
where $i=B,K$ (B-meson, kaon);  $f_{i}$, $F_{i}$ and $M_{i}$ are meson decay constants,
formfactors  and masses correspondingly (note that $F_B$ defined here correspond to $F_1$ from the previous subsection); $N$ is a numerical factor
identical for both $K$ and $B$-mesons. Therefore
\be
\left .\frac{(1/R)_{3}}{(1/R)_{2}}\right
|_B=\frac{m_B}{m_K}\lsb\frac{f_K}{f_B}\rsb^{1/2}\lsb\frac{F_B}{F_K}\rsb^{1/2}
\lsb\frac{Br_{3B}/Br_{2B}}{Br_{3K}/Br_{2K}}\rsb^{1/4} \left
.\frac{(1/R)_{3}}{(1/R)_{2}}\right |_K\;, \nn
\ee
The factors in square brackets are of order of  one (we assume that the BR are constrained on the 
same precision level separately for B and K-mesons).
So we see that due to the large ratio $m_B/m_K\simeq 11$
the three-body decay is more restrictive in the case of $B$-mesons.

\subsection{Processes with $\Delta G\neq 0$}
 \subsubsection{$B^0\to \mu e$}

 This decay has the strongest experimental constraint on BR among $\Delta
G\neq 0$ forbidden decays of $B^0$-meson. It is mediated by
$Z^\mu_{l,2}$-bosons carrying $G=2$ and $Z^\mu_{l,1}$ carrying $G=1$. But
vertex structure for $Z^\mu_{l,1}$ contains factor
$E^{l,1}_{12}-E^{l,1}_{23}$, what leads to an additional suppression of
order $\theta_A\simeq 0.1$ comparing to the $Z^\mu_{l,2}$
contribution~\cite{Frere:2003ye} and we omit it. So, the
corresponding interaction vertex is given by
\begin{equation}
  \G \sum_{l=2}^\infty
  Z^\mu_{l,2}E^{l,2}_{13}\left\{\bar{b}\gamma_\mu(-\frac12\gamma_5)d+
\epsilon_L\alpha_L\bar{e}
  \gamma_\mu(2\si-\frac12-\frac12\gamma_5)\mu\right\} \nn
\end{equation}
where $\epsilon_L\alpha_L\simeq0.13$ is small  parameter specifying
lepton mixing. An appearance of the factor $\epsilon_L\alpha_L$ in the
first power before the second term indicates that this term violates generation number (angular momentum) $G$
by unit. Successive treatment is just the same to that in the cases
of $B^0_s\to \mu e$ and $B^0\to \tau e$. The final result is the
following,
\begin{equation}
 \frac1R>m_Z\left(\frac{G^2_F\xi^2 (\epsilon_L \alpha_L)^2 f^2_{B^0}
m_{B^0}m^2_\mu\tau_{B^0} (1+(1-4\si)^2)}{64\pi B_{B^0\to \mu
e}}\right)^{1/4}\;, \nn
 \end{equation}
 where $B_{B^0\to \mu e}=6.4\times 10^{-8}$ is the experimental
 restriction on $B^0\to \mu e$ BR. Numerically,
 \begin{equation}
 \frac1R >0.15 \mbox{~TeV.} \nn
 \end{equation}
and for $R<64\mbox{~TeV}$:
\be
Br(B^{0}\to \mu e) \left |_{R<R_{K\to \mu e}}\right .<3.6\cdot 10^{-18}\;. \nn
\ee

\subsubsection{$B^0_s \to \mu^+\mu^- $}
This process is under keen interest of CMS and LHCb experiments~\cite{Adeva:2009ny}.
But unlike decays considered above it occurs in the SM through higher
order loop diagrams. SM prediction for its BR is
 $3.2\pm0.2\times10^{-9}$~\cite{Gamiz:2009ku, Buras:2010mh}. Current
 experimental limitation on $B^0_s \to \mu^+\mu^- $ BR is $B_{B^0_s \to
\mu^+\mu^-}=4.7\times10^{-8}$~\cite{Nakamura:2010zzi}. Due to the presence of
SM contribution further statistics accumulation will not lead to
significant improvement of a restriction, which we can obtain. Taking this
into account we will make rough estimation. Particularly, we will neglect
some cancellations that happen owing to the vertex structure.

$B^0_s \to \mu^+\mu^- $ decay occurs with $\Delta G=1$, so its amplitude
is suppressed by the first power of the mixing parameter $\eps_L\al_L$.
The corresponding vertex is given by
\be
\G\sum_{l=0}^\infty E_{2 3}^{l,1} Z_{l,1}^\mu \l -\h \ov{b}\g_\mu \g_5
s - \sqrt{2}\e_L\alpha_L \ov{\mu} \lsb \h\g_\mu \g_5 + \l \h - 2
\sin\theta_W \r \mu \rsb \r\;,  \nn
\ee
Performing calculations in the same manner as previously, we find,
 \begin{equation}
 \frac1R>m_Z\left(\frac{G^2_F\zeta^2 (\epsilon_L \alpha_L)^2 f^2_{B_s}
 m_{B_s}m^2_\mu\tau_{B_s} (1+(1-4\si)^2)}{8\pi B_{B^0_s \to
 \mu^+\mu^-}}\right)^{1/4}\, \nn
 \end{equation}
 and numerically,
 \begin{equation}
 \frac1R>0.5 \mbox{~TeV}\;. \nn
 \end{equation}

Experimental constraint on R, required  to obtain $1/R>64 ~\mbox{TeV}$ limit, is
\[
Br(B^0_s \to \mu^+\mu^-)|_{R<R_{K\to \mu e}}<1.6\cdot 10^{-16}\;,
 \]
and is negligible comparing  to the SM contribution.

 \subsubsection{$\Delta {m_{B_s}}$}
Gauge bosons carrying a non-zero
 angular momentum $G$ can contribute to the mass difference $\Delta m_B$.
 This mass difference appears due to the transitions $B^0_s\leftrightarrow
 \bar{B^0_s}$ with $\Delta G=2$. Corresponding contribution is
 \begin{equation}
 {\Delta}' m_B=2 Re \langle B^0_s|H_{\Delta G=2}|\bar{B^0_s}\rangle \nn
 \end{equation}
 and should be less than experimental value:  $\Delta m_{B_s}\approx
 1.17\cdot 10^{-8}$MeV.

Because of the large value of the strong interaction constant
$g_{s}\simeq1.1$, the dominant contribution to the
$\bar{b}s\leftrightarrow \bar{s}b$ transition originates from the gluon KK
modes exchange. The relevant interaction is
\eqsplit{&(\epsilon_d \gamma_d) g_s \sum\limits_{l=1}^\infty
\left(E_{22}^{l,0}-E_{33}^{l,0} \right) G_{l,0}^{\mu i} \bar b\gamma_\mu
{\lambda_i\over 2} s + {\rm h.c.}
 \label{Top}\nn}
Corresponding contribution to the mass difference
\begin{eqnarray*}
&\Delta 'm_{B}\approx 2\mbox{Re}\langle B_0|H_{\Delta
G=2}|\bar B_0 \rangle=&\nonumber\\
&\!\!\!\!\!\!\!\!= m_Bf_B^2
\displaystyle\frac{8g^2_{S}}{9}\left\{\!(\epsilon_d \gamma_d)^2
+(\epsilon_u \gamma_u)^2
+\left({m_B\over m_b+m_s}\right)^2\!\!\!\!
\epsilon_d \gamma_d\epsilon_u \gamma_u\!\right\}
\!\!\sum\limits_{l=1}^\infty
\left(E_{22}^{l,0}-E_{33}^{l,0}\right)^2 \!\!{R^2\over l(l+1)}\;,&
\end{eqnarray*}
where the matrix element was estimated by making use the vacuum insertion
approximation \cite{Beneke:1996gn}.


We note that, besides the expected $(\epsilon _{d}\alpha
_{d})^{2}$, there are two additional suppression factors
(see~\cite{Frere:2003ye}). First of them arises due to the structure of
the sum
\[
\sum_{l=0}^\infty (E^{l,0}_{22}-E^{l,0}_{33})^2\frac1{l(l+1)}\sim
{\theta_A}^2\;,
\]
where $\theta_A\simeq 0.1$. The second one is $\alpha _{u}\sim \delta
^{3}$ whereas $\alpha _{d}\sim \delta $. Therefore, for all parameters
taken from~\cite{Nakamura:2010zzi}, we obtain the limit on $R$,
\begin{eqnarray}
{1\over R}>(\epsilon_d \gamma_d) g_s f_B \theta_A \sqrt{\zeta {8\over 9}
\left(1+\left({m_B \over m_b+m_s}\right)^2\frac{\epsilon_u\gamma_u}
{\epsilon_d\gamma_d}\right)
\frac{m_B}{\Delta m_B}}\nonumber\\
 \simeq(\epsilon_d
\alpha_d\theta_A)\displaystyle\sqrt{\rule{0pt}{15pt}1+
1.6\displaystyle\frac{\epsilon_u\alpha_u}
{\epsilon_d\alpha_d}}\cdot90~{\rm TeV} \approx 90~{\rm GeV}.
\nonumber
\end{eqnarray}
We see that, if one even does not take into account all suppression
factors except $\alpha _{d}\sim \delta \sim 0.1$, the constraint on the
radius $R<(10 \mbox{ TeV})^{-1}$ will be less restrictive than one obtained from $K\to \mu e$ decay.

It is worth noting that in the model under consideration there are no
interactions which can contribute to $\Delta m_{B^0}$ since KK modes,
which are carrying angular momenta $G$ exceeding two units, do not interact
with the fermionic zero modes at tree level.

Higher excitations of gluon field could also contribute to the processes with $CP$ - violation, such as $B^0\to K^+\pi^-$ decay.
For the same reasons as for the mass difference, $CP$-violating processes are not restrictive enough, so we 
present only the final result:
\begin{equation*}
{1\over R} > (\epsilon_d\alpha_d\theta_A)^{1/2}\cdot 24\mbox{~TeV}\approx 0.75\mbox{~TeV}.
\end{equation*}

\section{Constraints from processes with D-mesons}
\label{Section/Pg16/1:bmesons/Constraints from processes with D-mesons}
To complete our consideration of rare heavy-meson decays we
shortly discuss processes with $D$-mesons, i.e. mesons that contain one
$c$-quark and one light quark.

We calculate the limits on $1/R$ from $D$-meson decay partial widths. One
can expect that constraints from $D$ decays would be less rigid than ones
from B decays. Really, the mass $m_D=1.9 \mbox{~GeV}$ three times smaller than
$m_B$. Also, mean lifetime of $D_0$ is four times smaller. Finally, experimental BR of
the forbidden $D$-meson decays are typically greater than BR in $B$-decays due to the low statistics.
Calculations similar to those performed in the previous sections lead to
$1/R<0.3 \mbox{~TeV}$ for $D_0 \to \mu e$ decay and this is the best
result from decays of $D$-mesons.

The decay $D_0 \to \mu^+\mu^-$ is caused by the box diagram in SM~\citep{PhysRevD.66.014009} 
Therefore, it is interesting from the experimental point of view. Using the
value $1/R=64 \mbox{~TeV}$ one can obtain BR of $D_0 \to \mu^+\mu^-$ in an
assumption that it is caused only by the heavy KK modes. This BR
is $Br(D\to\mu^+\mu^-)|_{R<R_{K\to \mu e}}<1.6\cdot
10^{-17}$, and so the contribution of the KK modes is
negligible as compared to SM one. Thus, decays of
$D$-mesons is less interesting in the context of
the model.

\section{Conclusions}
\label{Section/Pg17/1:bmesons/Conclusions}
In recent years, there has been considerable experimental progress in the
study of the physics of heavy $B$ and $D$-mesons. In this regard, we
returned to the question how rare or forbidden flavour-changing processes
may limit the six-dimensional model with a single generation of
vector-like fermions in the bulk ~\citep{Frere:2003ye, Libanov:2011zz,Libanov:2011st}. Previously, in the context of the model
it has been demonstrated how to explain an origin of the charged fermionic
generations of the SM fermions and fermionic mass hierarchy without
introducing a flavour quantum number: three families of four-dimensional
fermions appear as three sets of zero modes developed on a brane by a
single multi-dimensional family while the fermionic wave functions
inevitably produce a hierarchical mass matrix due to different overlaps
with the Higgs field profile.  In fact,  the role of a family number is
played by (almost) conserved angular momentum, corresponding to the rotations
in the extra dimensions, while the hierarchy is governed by one parameter
$\delta \sim 0.1$. It also has been shown, that massive neutrinos can be
easily incorporated into the model and have predicted an inverted
pseudo-Dirac mass pattern with $\Delta m_{12}^{2}/\Delta m_{13}^{2}\sim
\delta ^{2}\sim 0.01$ ~\citep{Frere:2010ah}, at least one maximal angle and one small
$\sin\theta _{13}\sim \delta \sim 0.1$ in the neutrino mixings
matrix\footnote{We would like to emphasize here, that, firstly, these
results are in a good agreement with the existing experemental
data~\cite{GonzalezGarcia:2010er, An:2012eh}, and, secondly, the parameter
$\delta $ in the neutrino sector is the \textit{same} as for the quark sector}.
We also have noted that higher excitations of the gauge bosons mediate
interesting neutral flavor-changing, but family-number conserving (in the
absence of mixings) interactions.  We have investigated ``usual''
flavour-changing processes, that are, as a rule, used to yield strongest
constraints on new physics. We have found, that the strongest limit on the
model arises from non-observation of the decay $K\to\mu e$; it requires
that the size of the extra-dimensional sphere (size of the gauge-boson
localization) $R$ satisfies $1/R \gtrsim 64$~TeV. A clear signature of the
model would be an observation of $K\to\mu e$ decay without  observations
of $\mu\to \bar{e}ee$, $\mu\to e\gamma$ and $\mu e$-conversion at the same
precision level~\citep{Frere:2003ye}.

In this paper we addressed specifically flavour-changing processes
involving $B$ and $D$-mesons with the aim to single out those processes
that yield the strongest constraint on the size $R$. We found that
the best limit $1/R>3.3\mbox{ TeV}$ arises from the three-body
decay $B^{0}\to K\mu e$ in contrast to the two-body decay $K\to
\mu e$ in kaons. This bound is much less stringent than the constraint
arising from $K\to \mu e$. The reason is, of course, in the still too poor
statistics: the experimental bound on the branching ratio of $K\to \mu e$
is $2.4\cdot 10^{-12}$ while for the $B$-meson decay is $2.7\cdot
10^{-7}$. However, it is hoped that thanks to the current and future
experiments statistics will be improved. In particular, to reach the
kaonic constraint on $1/R \gtrsim 64$~TeV one needs to limit the $B$-meson
branching ratio on the level $2.4\cdot 10^{-12}$. Interestingly, that this
is the same level as for kaons. It means in particular, that the
distinctive feature of the model would be an observation $K\to\mu e$
\textit{and} $B^{0}\to K\mu e$ decays without  observations other
flavour-changing processes at the same precision level.

\section*{Acknowledgments} We are indepted to J.-M.~Frere, D. Gorbunov and
S.~Troitsky for helpful discussions. We esecailly thank D.~Gorbunov for stimulating
this work. This work is supported in part by the grant of the President of
the Russian Federation NS-5590.2012.2; by Russian Foundation for Basic
Research grants 12-02-00653, 11-02-92108 (M.L.); by fellowships of the
``Dynasty'' foundation (M.L. and I.T.).

\bibliography{3gen}
\bibliographystyle{JHEP}
\end{document}